# A Comparative Study of Multipole and Empirical Relations Methods for Effective Index and Dispersion Calculations of Silica-Based Photonic Crystal Fibers

Ashkan Ghanbari[a], Alireza Kashaninia[a], Ali Sadr[b], and Hamed Saghaei[c]

[a]Department of Electrical Engineering, Faculty of engineering, Central Tehran Branch, Islamic Azad University, Tehran, Iran
[b]Faculty of Electrical and Electronics Engineering, Iran University of Science and Technology, Tehran, Iran.
[c]Department of Electrical Engineering, Shahrekord Branch, Islamic Azad University, Shahrekord, Iran.
ashkan.ghanbari@iauctb.ac.ir, ali.kashaniniya@iauctb.ac.ir, sadr@iust.ac.ir, h.saghaei@iaushk.ac.ir
Corresponding author: Alireza Kashaninia

**Abstract-** In this paper, we present a solid-core Silica-based photonic crystal fiber (PCF) composed of hexagonal lattice of air-holes and calculate the effective index and chromatic dispersion of PCF for different physical parameters using the empirical relations method (ERM). These results are compared with the data obtained from the conventional multipole method (MPM). Our simulation results reveal that the ERM is an accurate and fast method for dispersion analysis of PCFs with large pitch sizes. However, for small pitch sizes of PCFs, it is not as accurate as the MPM method. Therefore, ERM is a fast, simple and accurate method for modelling and analysis of Silica-based PCFs with large pitch sizes.

*Index Terms-* Dispersion, empirical relations method, multipole method, photonic crystal fiber, Silica.

I. INTRODUCTION

Photonic crystal fiber (PCF) is a kind of optical fiber that uses photonic crystals to form the cladding around the core of the cable. Photonic crystal is a low-loss periodic dielectric medium constructed using a periodic array of microscopic air holes that run along the entire fiber length.Many resarchs were done in this field of science.For example, zahedi et al Designed and simulated an Optical Demultiplexer Using Photonic Crystals in 2017 [1] or many logic gates and logical circuits were designed and implemented based on photonic crystals such as XOR,NOR, decoder , encoder and etc [2-4]by use of several methods [2-4]. Photonic crystal fibers (PCFs) as a new class of optical waveguides have been studied by many research groups due to their wide applications in science and technology [5-8]. The existence of various lattice structures of air holes such as circular [6], square [9], and hexagonal [10, 11] in the cladding area of PCFs leads to the appearance of many unique features such as dispersion engineering [12, 13], high or low nonlinearity [14], adjustable zero dispersion wavelengths [15] compared to the conventional optical fibers.

Due to the small values of cladding index compared to the solid core of PCFs that usually made of Silica, the guiding mechanism is provided by the total internal reflection (TIR) along the solid core of the fiber [16, 17]. PCFs can be made of various materials such as Silica [14], Silicon [19-21], Chalcogenide [15, 22-25], Fluoride [26], and Tellurite [27]. Many theoretical and experimental efforts are still being pursued by the researchers and scientists to analyze the propagation characteristics of PCFs made of different materials using various methods . There are several techniques to study the PCF characteristics including: multipole method (MPM) [28, 29], finite element method (FEM) [30], finite difference method eigenmode (FDE) [31], plane wave expansion method (PWM) [32] and effective index method (EIM) [33]. Although they are very high accuracy methods, they have a lot of time-consuming, and difficult calculations. There are several easier analytical and empirical methods such as vectorial effective index method, scalar effective index method [34] and empirical relations method (ERM) [35]. They are more cost effective than the numerical techniques. Several reports were published about the propagation characteristics analysis of Silica-based PCFs using the ERM method up to now [35-37]. Furthermore, there are some reports in which the accuracy of EIM was compared with ERM that the results indicated ERM has more accuracy in comparison with EIM. Due to the researchers' tendency to provide easier and time-saving methods for analysis and design of propagation characteristics of PCFs, a comprehensive study of these methods is required. To the best of our knowledge, such analysis has not been studied so far. In this report, the linear profiles of a Silica-based solid-core PCF for different values of physical parameters are calculated using the ERM. The results are compared with the data achieved by MPM as a well-known and accurate method to define the exact functional ranges of ERM. Totally this article includes the following points:

• The ERM and MPM accuracy Comparison for dispersion analysing of the Silica photonic crystal fibers covering visible to NIR wavelengths.
• Defining the exact functional ranges of empirical relation method in dispersion and effective index analysizng for the Silica PCFs.

II.  PCF PHYSICAL STRUCTURE

Fig. 1 illustrates the cross-sectional view of solid core Silica-based PCF, under study, made by devising circular air-holes (white circles) of radii $d$ in a Silica background. The PCF's six circular air-holes that are arranged in a hexagonal lattice of pitch size $\Lambda$, encompassing the solid core of diameter $d_c=2\Lambda-d$.

## III. NUMERICAL METHODS

### A. Empirical Relations Method

PCF parameterization using ERM in terms of the V is calculated as follows [11, 32, 35]:

$$\underbrace{\frac{2\pi a_{eff}}{\lambda}\sqrt{n_{co}^2-n_{FSM}^2}}_{V} = \sqrt{(\underbrace{\frac{2\pi a_{eff}}{\lambda}\sqrt{n_{co}^2-n_{eff}^2}}_{U})^2 + (\underbrace{\frac{2\pi a_{eff}}{\lambda}\sqrt{n_{eff}^2-n_{FSM}^2}}_{W})^2} \quad (1)$$

where $a_{eff}$ is the core radius ($a_{eff}=\Lambda/\sqrt{3}$). $n_{co}$ is the core refractive index. It is supposed to be 1.45 for Silica at 1.55μm [10]. V and W are normalized transverse and attenuation terms, respectively. $n_{fsm}$ is the cladding index. $n_{eff}$ is the effective refractive index that is a key parameter of every PCF characterization profile. It is related to the propagation constant, $\beta$, through $\beta=n_{eff}\times k$ where $k=2\pi/\lambda$ is wave number in free space and finally λ is wavelength of optical beam. From Ref. [35], we realized that the V parameter of Silica-based hexagonal PCFs can be estimated as equation below [28]. It is in terms of normalized hole diameter of $d/\Lambda$ and normalized wavelength of $\lambda/\Lambda$:

$$V\left(\frac{d}{\Lambda},\frac{\lambda}{\Lambda}\right) = \frac{2\pi a_{eff}}{\lambda}\sqrt{n_{co}^2-n_{FSM}^2} = \frac{A_2}{1+A_3 e^{(A_4\frac{\lambda}{\Lambda})}} + A_1 \quad (2)$$

where $A_i(i=1,..,4)$ are fitting parameters and can be derived from trial and error method for different values of $d/\Lambda$ in a wide range of wavelengths by considering Ref. [35]. As a sample, in this study, the fitting parameters of $d/\Lambda=0.8$, 0.6, and 0.4 and $\Lambda$=850nm for a central wavelength of 800nm are presented in Table I. By calculating the desired $V$ parameter of the fiber, the values of the cladding index can be found through Eq. (1) as follows [35]:

$$n_{FSM} = \sqrt{(1.45)^2 - V(d/\Lambda,\lambda/\Lambda)^2 \lambda^2/12\Lambda^2} \quad (3)$$

The attenuation constant ($W$) of the fiber for different values of $d/\Lambda$ in a wide range of wavelengths can be computed as follows [35]:

$$W\left(\frac{d}{\Lambda},\frac{\lambda}{\Lambda}\right) = \frac{2\pi a_{eff}}{\lambda}\sqrt{n_{eff}^2-n_{FSM}^2} = \frac{B_2}{1+B_3 e^{(B_4\frac{\lambda}{\Lambda})}} + B_1 \quad (4)$$

where $B_i(i=1,2,3,4)$ are fitting parameters which are related to $W$ parameter.

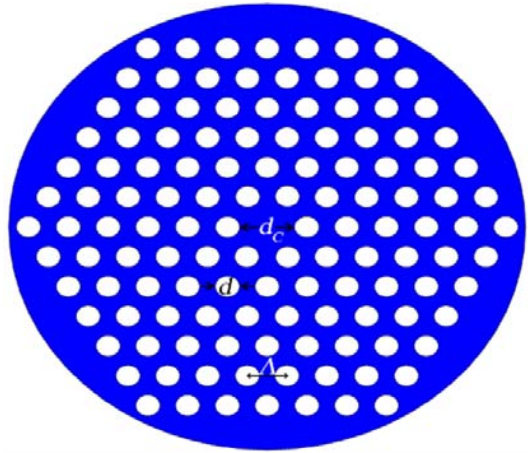

Fig. 1. Cross-sectional view of solid core PCF with six hexagonal rings of air-holes.

TABLE I. FITTING PARAMETERS OF EQ. (2)

| Parameters | $A_1$ | $A_2$ | $A_3$ | $A_4$ |
|---|---|---|---|---|
| $d/\Lambda = 0.8$ | 1.1045 | 13.7830 | 1.6108 | 1.4947 |
| $d/\Lambda = 0.6$ | 0.7281 | 5.7153 | 1.0300 | 1.2638 |
| $d/\Lambda = 0.4$ | 0.5844 | 2.6392 | 0.5622 | 1.5363 |

TABLE II. FITTING PARAMETERS OF EQ. (4)

| Parameters | $B_1$ | $B_2$ | $B_3$ | $B_4$ |
|---|---|---|---|---|
| $d/\Lambda = 0.8$ | 0.0143 | 15.1949 | 1.5343 | 1.3731 |
| $d/\Lambda = 0.6$ | -0.1922 | 5.4698 | 0.6930 | 1.3662 |
| $d/\Lambda = 0.4$ | -0.1123 | 2.3700 | 0.3301 | 2.0709 |

As a sample, In this study, the fitting parameters of $d/\Lambda = 0.8$, 0.6, and 0.4 and $\Lambda$=850nm for a central wavelength of 800nm are presented in Table II. From Refs .[34] and [35], we realized that these parameters can be calculated for different values of $d/\Lambda$ and wavelengths. Consequently, using Eq. (3) for given cladding index and *W*, the effective refractive index of our PCF for $d/\Lambda = 0.8$ ,0.6 and 0.4 at different wavelengths can be achieved.

The chromatic (total) dispersion of the proposed PCF is calculated as follows:

$$D_t(\lambda) = -\frac{\lambda}{c}\frac{d^2}{d\lambda^2}\text{Re}\left[n_{\text{eff}}(\lambda)\right] \tag{5}$$

*B. Multipole Method*

In multipole method (MPM), each of the air holes that is embedded in the cladding region of the PCFs plays the role of scattering cell. The electromagnetic field can be described by the Henkel function in the cylindrical system. Thus, boundary conditions can be the solutions of the Helmholtz equation. The description of the electric field in the z-direction of the nth air holes of the PCF can be defined as:

$$E_Z = \sum_{m=-\infty}^{\infty} a_m(n) J_m(k_\perp^i r_n) \exp(i_m \phi_n) \exp(i\beta z) \tag{6}$$

The electric field around $n^{th}$ air hole inside the Silica background is given by:

$$E_Z = \sum_{m=-\infty}^{\infty} b^n{}_m J_m(k_\perp^i r_n) \exp(i_m \phi_n) \exp(i\beta z) + C^n{}_m H^n{}_m(k_\perp^e r_n) \exp(i_m \phi_n) \exp(i\beta z) \tag{7}$$

where

$$k_\perp^i = (k_0^2 n_i^2 - \beta^2)^{0.5} \quad k_\perp^e = (k_0^2 n_e^2 - \beta^2)^{0.5} \tag{8}$$

Where $n_i$ and $n_e$ are the refractive indices of the air and Silica, respectively. As mentioned earlier, $k$ parameter is responsible for the wavenumber in free space and $\beta$ stands for propagation mode constant. The components of electric and magnetic fields have the same definitions, thus, in this case using boundary conditions, the coefficients of $a_m, b_m, c_m$ can be found. By calculation of the mentioned coefficients and using $\beta = n_{eff} \times k$, the effective refractive of the related PCF can be accurately defined. In this paper, the most widely used CUDMOF software based on Multipole method is applied to analyze the related optical properties of the fiber rigorously.

IV. RESULTS AND DISCUSSION

Our simulation results compare the effective refractive index and total dispersion of the proposed Silica-based PCF for $\Lambda = 0.85$, 1 and 2μm with a fixed normalized air-hole diameter (air-hole diamater to pitch ratio) of $d/\Lambda = 0.8$ using MPM and ERM. Fig. 2 depict the calculated effective refractive index, $n_{eff}$, including two-period moving average lines based on ERM and MPM. Trend line analysis of these figures clearly show that the deviation of the lines increases when the pitch size of the fiber decreases. The smaller pitch size of the PCF results in more deviation of the effective refractive index. It means that the mentioned deviations make more difference when the design is based on smaller pitches (especially for $\Lambda < 1\,\mu m$). It is because of the fitting parameter deviations which occur in smaller pitch sizes. But the accuracy of ERM method becomes closer to MPM using pitches more than 1μm. For instance, PCF with $d/\Lambda = 0.8$, $\Lambda = 0.85\,\mu m$ and the central wavelength of 800nm, the relative difference between ERM and MPM (ERM-MPM) for the effective index is 0.016 and it becomes 0.00029 and zero for $\Lambda = 1$ and $2\mu m$ respectively.

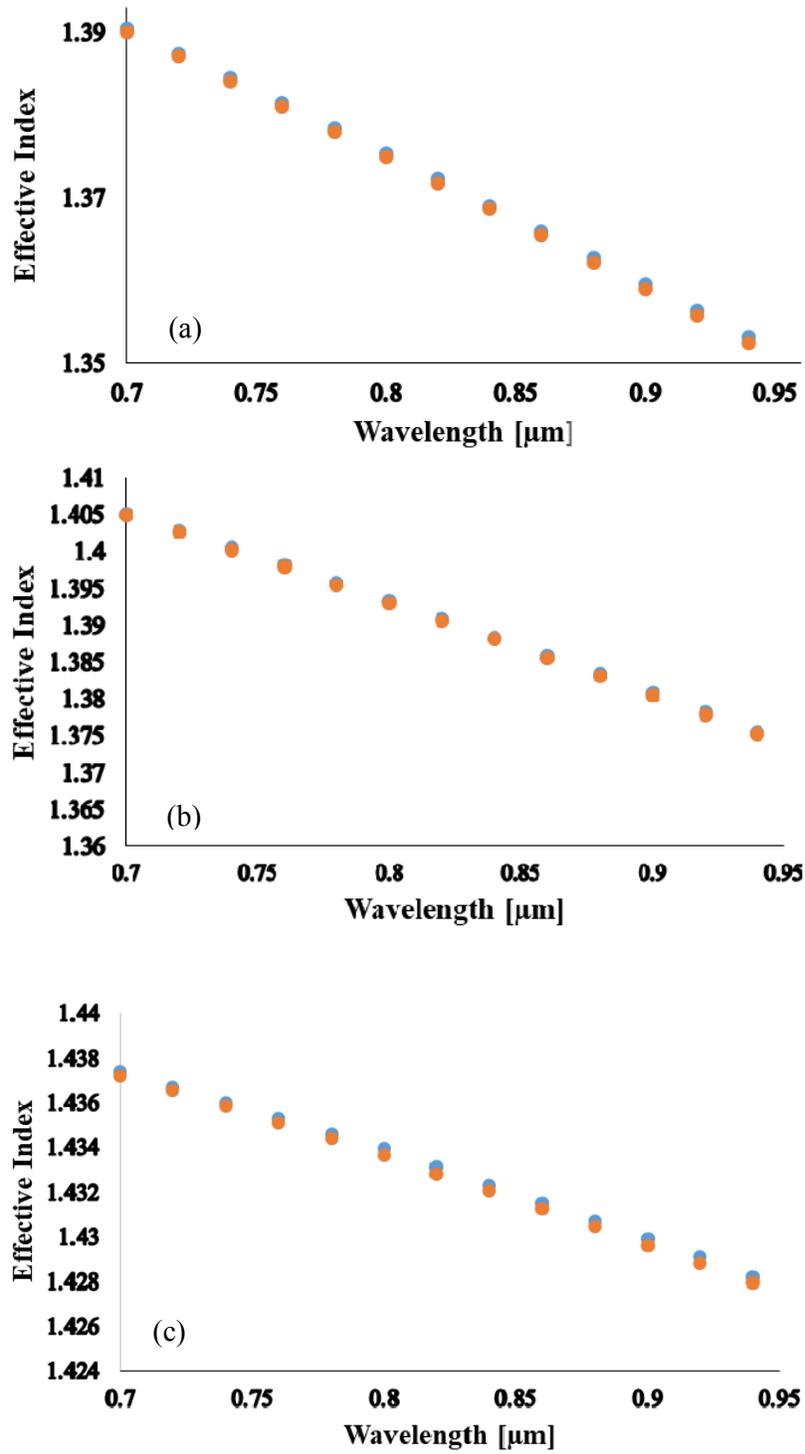

Fig. 2. The effective refractive index of the proposed Silical-based PCF for $d/\Lambda = 0.8$ and (a) $\Lambda = 0.85$ μm, (b) $\Lambda = 1$ μm, and (c) $\Lambda = 2$ all using MPM (orange circles) and ERM (blue circles).

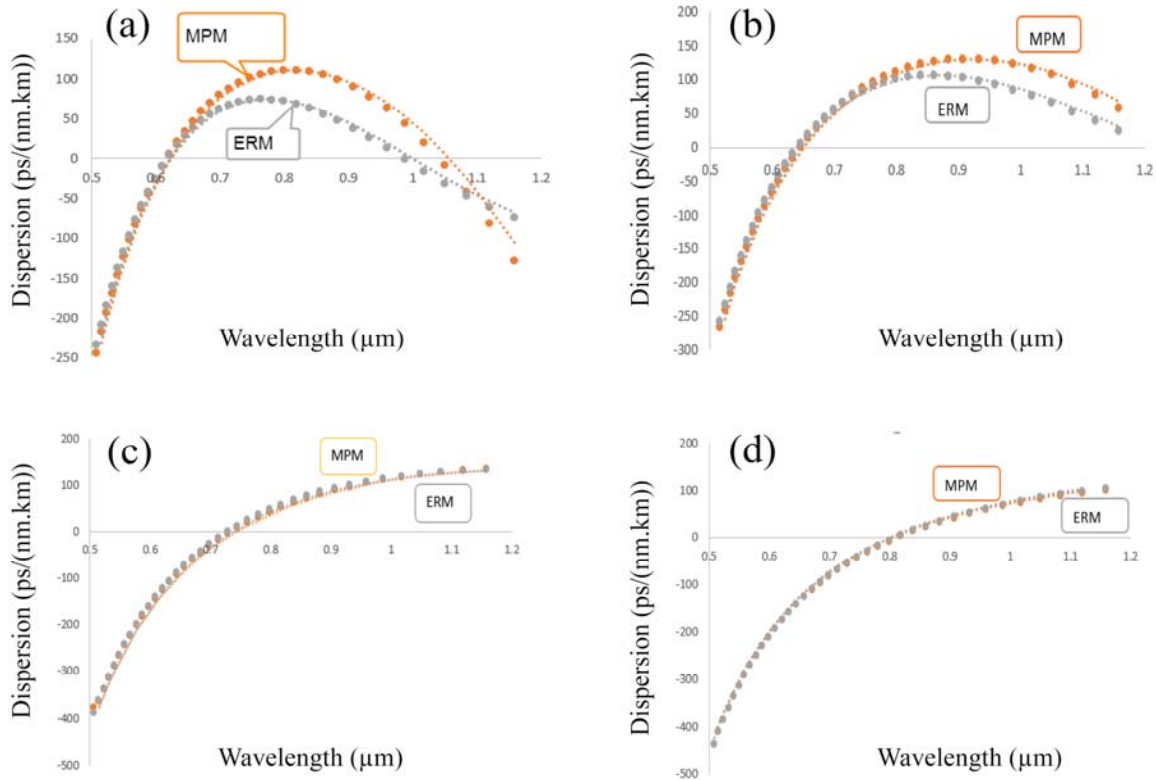

Fig. 3. Total dispersion of the proposed Silica-based PCF for $d/\Lambda = 0.8$ and (a) $\Lambda = 0.85\,\mu m$, (b) $\Lambda = 1\,\mu m$, (c) $\Lambda = 1.5\,\mu m$ and (d) $\Lambda = 2\,\mu m$ all using MPM (orange circles) and ERM (gray circles).

It should be noted a small error in the calculations of the effective refractive index results in the difference between other parameters such as total dispersion, higher order dispersions. The chromatic dispersion of the fundamental mode using ERM and MPM are shown in Fig. 3. In these figures, we can see a perceptible decrease in the difference between these two methods by increasing the pitch size. As expected, the accuracy of the dispersion of both methods are not acceptable at smaller values of pitch sizes (specifically for $\Lambda < 1\,\mu m$). For instance, at wavelength of 850nm the calculated value for total dispersion based on MPM is 104.68 ps/km.nm and based on ERM is 55.33(ps/km.nm) for PCF with $d/\Lambda = 0.8$ and $\Lambda = 0.85\,\mu m$ which shows a large difference of 49.34 (ps/km.nm). Therefore, ERM accuracy cannot be acceptable for an accurate design in this range of pitch sizes. While by choosing $\Lambda = 1$, 1.5 and also $2.5\mu m$, the values of total dispersion change to 106.5768 (ps/km.nm), 76.6073 (ps/km.nm) and 25.4183 (ps/km.nm) based on ERM and 126.809 (ps/km.nm), 71.296 (ps/km.nm) and 23.0689 (ps/km.nm) based on MPM which show acceptable relative difference values of 20.23 (ps/km.nm), -5.311 (ps/km.nm) and -2.34 (ps/km.nm), respectively. We repeated our calculations for $d/\Lambda = 0.6$, 0.4 and $\Lambda = 0.85$, 1 and $1.5\mu m$. The results are shown in Fig. 4. Comparing the results clearly show that, the pitch size plays an important role on the analysis of the dispersion. Also, it is obvious that, even for different values of pitch sizes and normalized air hole diameters the difference between the accuracy of MPM and ERM increase with decreasing of the pitch sizes for $\Lambda < 1$ μm. For example, at the central wavelength of 850nm and structure parameters of $d/\Lambda = 0.6$ and $\Lambda = 0.85$ μm, the calculated values of the total dispersion is -121.5 (ps/km.nm) based on MPM while this value changes to -62.421(ps/km.nm) based on ERM which shows a large difference of 59.133(ERM-MPM) or in another case, for structure parameters of $d/\Lambda = 0.4$ and $\Lambda = 1.5$ μm, the calculated values of the total dispersion is -79.9(ps/km.nm) based on MPM while this value changes to -88.2879(ps/km.nm) based on ERM at 850nm which shows a negligible difference of 8.3(ps/km.nm). Finally, we can present the value of $\Lambda = 1\,\mu m$ as the border point of ERM accuracy in which by increasing the values of pitch sizes over than it, the accuracy starts to approach MPM and by decreasing the values of pitch sizes less than 1μm the accuracy begins to decrease from the MPM.

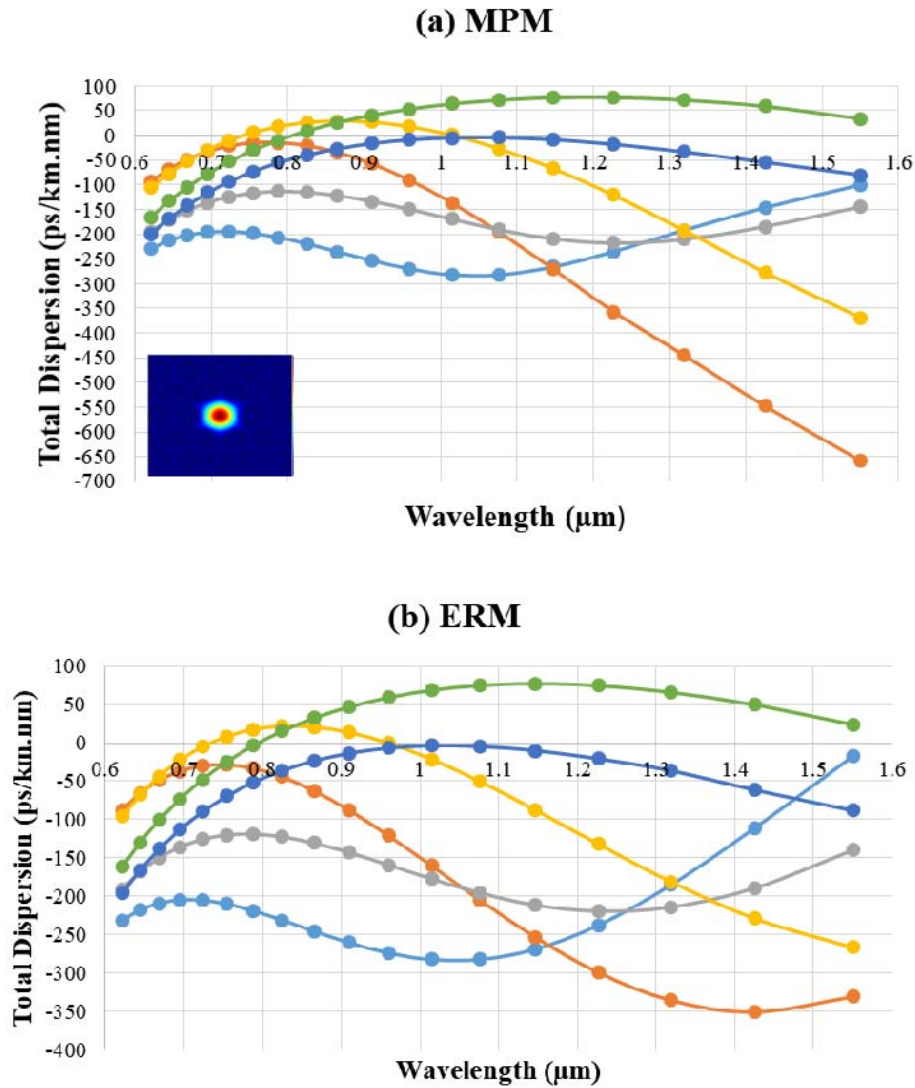

Fig. 4. Total dispersion of Silica-based PCF based on (a) ERM, and (b) MPM. The inset of Fig. 4(a) shows 2D view of fundamental mode distribution inside the PCF. Green line is for Λ=1.5μm, *d*=0.9μm, dark blue line is for Λ=1.5μm, *d*=0.6μm, yellow line is for Λ=1μm, d=0.6μm, orange line is for Λ=1μm, *d*=0.4μm, gray line is for Λ=0.85μm, *d*=0.51μm and blue line is for Λ=0.85μm, and *d*=0.34μm.

## V.  CONCLUSION

In summary, the linear parameters of a Silica-based solid-core PCF including effective refractive index and chromatic dispersion were studied using empirical relations method (ERM) and compared with multipole method (MPM). We demonstrated that although ERM is an easy and fast method, its application is limited to PCFs with pitch sizes less than 1μm and studying the propagation characteristics of light in this range of pitch sizes faces our designed PCF with challenges. Our numerical calculations illustrated that ERM's accuracy becomes closer to MPM for PCF with pitch sizes more than 1μm. This accuracy is even much closer to MPM for the PCF with pitch sizes more than 1.5μm. Consequently, this method is recommended to be used for modelling and analysis of Silica-based PCFs with large pitch sizes because of its simplicity and high speed.

REFERENCES

[1] A. Zahedi, F. Parandin, M. M. Karkhanehchi, H. Habibi Shams, and S. Rajamand "Design and Simulation of Optical 4-Channel Demultiplexer Using Photonic Crystals," *Journal of Optical Communications*, vol.2, no.5, pp. 22-25, May 2013.

[2] M. M. Karkhanehchi, F. Parandin, A. Zahedi, "Design of an all optical half-adder based on 2D photonic crystals," *Photonic Network Communications*, vol. 33, no. 2, pp. 159-165, April 2017.

[3] F. Parandin, R. Malmir, M. Naseri, and A. Zahedi, "Reconfigurable all-optical NOT, XOR, and NOR logic gates based on two dimensional photonic crystals," *Superlattices Microstruct.*, vol.113, no. 2018, pp. 737-744 , January 2018.

[4] F. Parandin, M. Mehdi Karkhanehchi, M. Naseri, and A. Zahedi, "Design of a high bitrate optical decoder based on photonic crystals," *Journal of Computational Electronics*, vol. 17, no. 2, pp.830-836, March 2018.

[5] A. Ghanbari, A. Kashaninia, A. Sadr, and H. Saghaei, "Supercontinuum generation for optical coherence tomography using magnesium fluoride photonic crystal fiber," *Optik-International J. for Light. Electron Opt.*, vol. 140, pp. 545-554, July 2017.

[6] H. Saghaei, and A. Ghanbari, "White light generation using photonic crystal fiber with sub-micron circular lattice," *J. Electrical Eng.*, vol. 68, no. 4, pp. 282-289, August 2017.

[7] H. Saghaei, B. Seyfe, H. Bakhshi, and R. Bayat, "Novel approach to adjust the step size for closed-loop power control in wireless cellular code division multiple access systems under flat fading," *IET Commun.*, vol. 5, no. 11, pp. 1469-1483, July 2011.

[8] H. Saghaei, and B. Seyfe, "New Approach to Closed-Loop Power Control in Cellular CDMA Systems under Multipath Fading," Proc. *IEEE Wicom'08 Conf.*, pp. 1-4, 2008,

[9] H. Saghaei, "Supercontinuum source for dense wavelength division multiplexing in square photonic crystal fiber via fluidic infiltration approach," *Radioengineering*, vol. 26, no. 1, pp. 16-22, March 2017.

[10] M. Ebnali-Heidari, H. Saghaei, F. Koohi-Kamali, M. N. Moghadasi, and M. K. Moravvej-Farshi, "Proposal for supercontinuum generation by optofluidic infiltrated photonic crystal fibers," *IEEE J. Sel. Top. Quantum Electron.*, vol. 20, no. 5, pp. 582-589, March 2014.

[11] H. Saghaei, V. Heidari, M. Ebnali-Heidari, and M. R. Yazdani, "A systematic study of linear and nonlinear properties of photonic crystal fibers," *Optik-International J. for Light. Electron Opt.*, vol. 127, no. 24, pp. 11938-11947, March 2016.

[12] M. Ebnali-Heidari, F. Dehghan, H. Saghaei, F. Koohi-Kamali, and M. K. Moravvej-Farshi, "Dispersion engineering of photonic crystal fibers by means of fluidic infiltration," *J. Mod. Opt.*, vol. 59, no. 16, pp. 1384-1390, August 2012.

[13] A. Ghanbari, A. Sadr, and H. Hesari, "Modeling photonic crystal fiber for efficient soliton-effect compression of femtosecond optical pulses at 850 nm," *Arab. J. Sci. Eng.*, vol. 39, no. 5, pp. 3917-3923, May 2014.

[14] A. Ghanbari, A. Karimkhani nia, and A. Sadr, "Superlattice elliptical-core photonic crystal fiber soliton effect compressor at 1550nm," *Journal of Communication Engineering.*, vol. 4, no.1, pp. 29-40, January-June 2015.

[15] H. Saghaei, M. K. Moravvej-Farshi, M. Ebnali-Heidari, and M. N. Moghadasi, "Ultra-wide mid-infrared supercontinuum generation in As 40 Se 60 chalcogenide fibers: solid core PCF versus SIF," *IEEE J. Sel. Top. Quantum Electron.*, vol. 22, no. 2, pp. 279-286, September 2016.

[16] B. Tan, X.-w. Chen, and S.-C. Li, "Total Internal Reflection Photonic Crystal Fiber," *J. Optoelectronics Laser,* vol. 13, no. 5, pp. 491-495, July 2002.

[17] H. Saghaei, A. Zahedi, R. Karimzadeh, and F. Parandin, "Line defects on As2Se3-Chalcogenide photonic crystals for the design of all-optical power splitters and digital logic gates," *Superlattices Microstruct.*, vol. 110, no.7, pp. 133-138, August 2017.

[18] J. Knight, J. Arriaga, T. Birks, A. Ortigosa-Blanch, W. Wadsworth, and P. Russell, "Anomalous dispersion in photonic crystal fiber," *IEEE Photon. Technol. Lett.*, vol. 12, no. 7, pp. 807-809, August 2000.

[19] M. Ebnali-Heidari, H. Saghaei, C. Monat, and C. Grillet, "Four-wave Mixing Based Mid-span Phase Conjugation using slow light engineered Chalcogenide and silicon photonic crystal waveguides," The European *Conference on Lasers and Electro-Optics*, pp. CD4_3, May 2011.

[20] J. Rarity, J. Fulconis, J. Duligall, W. Wadsworth, and P. S. Russell, "Photonic crystal fiber source of correlated photon pairs," *Opt. Express*, vol. 13, no. 2, pp. 534-544, July 2005.

[21] H. Saghaei and V. Van, "Broadband mid-infrared supercontinuum generation in dispersion-engineered silicon-on-insulator waveguide," *JOSA B,* vol. 36, no. 2, pp. 193-202, February 2019.

[22] M. Diouf, A. B. Salem, R. Cherif, H. Saghaei, and A. Wague, "Super-flat coherent supercontinuum source in As 38.8 Se 61.2 chalcogenide photonic crystal fiber with all-normal dispersion engineering at a very low input energy," *Appl. Opt.*, vol. 56, no. 2, pp. 163-169, January 2017.

[23] H. Saghaei, "Dispersion-engineered microstructured optical fiber for mid-infrared supercontinuum generation," *Appl. Opt.*, vol. 57, no. 20, pp. 5591-5598, July 2018.

[24] R. Raei, M. Ebnali-Heidari, and H. Saghaei, "Supercontinuum generation in organic liquid-liquid core-cladding photonic crystal fiber in visible and near-infrared regions," *JOSA B*, vol. 35, no. 2, pp. 323-330, Febrarury 2018.

[25] H. Saghaei, M. Ebnali-Heidari, and M. K. Moravvej-Farshi, "Midinfrared supercontinuum generation via As 2 Se 3 chalcogenide photonic crystal fibers," *Appl. Opt.*, vol. 54, no. 8, pp. 2072-2079, March 2015.

[26] G. Qin, X. Yan, C. Kito, M. Liao, C. Chaudhari, T. Suzuki, and Y. Ohishi, "Ultrabroadband supercontinuum generation from ultraviolet to 6.28 µ m in a fluoride fiber," *Appl. Phys. Express,* vol. 95, no. 16, pp. 161103, October 2009.

[27] V. R. K. Kumar, A. George, J. Knight, and P. Russell, "Tellurite photonic crystal fiber," *Opt. Express*, vol. 11, no. 20, pp. 2641-2645, November 2003.

[28] B. T. Kuhlmey, T. P. White, G. Renversez, D. Maystre, L. C. Botten, C. M. de Sterke, and R. C. McPhedran, "Multipole method for microstructured optical fibers. II. Implementation and results," *JOSA B,* vol. 19, no. 10, pp. 2331-2340, August 2002.

[29] T. White, B. Kuhlmey, R. McPhedran, D. Maystre, G. Renversez, C. M. De Sterke, and L. Botten, "Multipole method for microstructured optical fibers. I. Formulation," *JOSA B,* vol. 19, no. 10, pp. 2322-2330, July 2002.

[30] F. Bréchet, J. Marcou, D. Pagnoux, and P. Roy, "Complete analysis of the characteristics of propagation into photonic crystal fibers, by the finite element method," *Opt. Fiber Technol.*, vol. 6, no. 2, pp. 181-191, April 2000.

[31] S. Guo, F. Wu, S. Albin, H. Tai, and R. Rogowski, "Loss and dispersion analysis of microstructured fibers by finite-difference method," *Opt. Express,* vol. 12, no. 15, pp. 3341-3352, July 2004.

[32] Y. Cao, Z. Hou, and Y. Liu, "Convergence problem of plane-wave expansion method for phononic crystals," *Phys. Lett. A,* vol. 327, no. 2, pp. 247-253, June 2004.

[33] M. Qiu, "Effective index method for heterostructure-slab-waveguide-based two-dimensional photonic crystals," *Appl. Phys. Lett.*, vol. 81, no. 7, pp. 1163-1165, June 2002.

[34] R. Sinha, and A. Varshney, "Dispersion properties of photonic crystal fiber: comparison by scalar and fully vectorial effective index methods," *Opt. Quantum Electron.,* vol. 37, no. 8, pp. 711-722, June 2005.

[35] K. Saitoh, and M. Koshiba, "Empirical relations for simple design of photonic crystal fibers," *Opt. Express,* vol. 13, no. 1, pp. 267-274, January 2005.

[36] M. D. Nielsen, N. A. Mortensen, M. Albertsen, J. R. Folkenberg, A. Bjarklev, and D. Bonacinni, "Predicting macrobending loss for large-mode area photonic crystal fibers," *Opt. Express,* vol. 12, no. 8, pp. 1775-1779, May 2004.

[37] S. Bandyopadhyay, P. Biswas, A. Pal, S. K. Bhadra, and K. Dasgupta, "Empirical relations for design of linear edge filters using apodized linearly chirped fiber Bragg grating," *J. Lightwave Technol.*, vol. 26, no. 24, pp. 3853-3859, Dec. 2008.